\input{aipcheck}

\documentclass[
    ,final            
  ]
  {aipproc}

\layoutstyle{6x9}

\begin{document}
\def\arcmin{\hbox{$^\prime$}}
\def\arcsec{\hbox{$^{\prime\prime}$}}
\def\degree{$^{\circ}$} 

\def\saxj{SAX~J1808.4--3658}
\def\igr{IGR~J00291+5934}
\def\1814{XTE~J1814--338}
\def\0929{XTE~J0929--314}
\def\lta{\la}
\def\gta{\ga}
\def\msun{{$\rm M_{\odot}$}}
\def\Porb{P_{\rm orb}}
\def\Pdot{\dot{P}_{\rm orb}}
\def\Pspin{P_{\rm spin}}
\def\L{{\cal L}}
\def\T0{T^*_0}
\def\asini{a_1 \sin i}

\def\ie{{\frenchspacing\it i.e. }}
\def\eg{{\frenchspacing\it e.g. }}
\def\etal{{~et al.~}}

\title{Order in the chaos? The strange case of accreting millisecond pulsars}

\author{Tiziana Di Salvo}{
  address={Dipartimento di Scienze Fisiche ed Astronomiche, Universit\`a di 
  Palermo, via Archirafi 36 - 90123 Palermo, Italy} }
\author{Luciano Burderi}{
  address={Universit\`a degli Studi di Cagliari, Dipartimento
  di Fisica, SP Monserrato-Sestu, KM 0.7, 09042 Monserrato, Italy} }
\author{Alessandro Riggio}{
  address={Universit\`a degli Studi di Cagliari, Dipartimento
  di Fisica, SP Monserrato-Sestu, KM 0.7, 09042 Monserrato, Italy} }
\author{Alessandro Papitto}{
  address={Dipartimento di Fisica, Universitá degli Studi di Roma `Tor Vergata', via
  della Ricerca Scientifica 1, 00133 Roma, Italy}, altaddress={Osservatorio 
  Astronomico di Roma, Via Frascati 33, 00040 Monteporzio Catone (Roma), Italy } } 
\author{Maria Teresa Menna}{address={Osservatorio Astronomico di Roma, 
  Via Frascati 33, 00040 Monteporzio Catone (Roma), Italy } }

\begin{abstract}
We review recent results from the X-ray timing of accreting
millisecond pulsars in LMXBs. This is the first time a timing analysis
is performed on accreting millisecond pulsars, and for the first time
we can obtain information on the behavior of a very fast pulsar
subject to accretion torques. We find both spin-up and spin-down
behaviors, from which, using available models for the accretion
torques, we derive information on the mass accretion rate and magnetic
field of the neutron star in these systems. We also find that the phase
delays behavior as a function of time in these sources is sometimes
quite complex and difficult to interpret, since phase shifts, most
probably driven by variations of the X-ray flux, are sometimes
present.
\end{abstract}

\maketitle


\section{Introduction}

According to the recycling scenario (see e.g. \citep{bha91} for a review), there exists an evolutionary connection
between the so-called Low Mass X-ray Binaries (LMXBs) containing a
neutron star and Millisecond Radio Pulsars (MSP). The first class of
sources consists of old systems where a low magnetized ($\sim 10^8 -
10^9$ Gauss) neutron star accretes matter from a low mass (usually
less or of the order of \msun) companion.  The weak magnetic field of
the neutron star allows the matter to be accreted very close to the
compact object; the accretion radius is indeed the magnetospheric
radius (the radius at which the magnetic pressure due to the assumed
dipolar magnetic field of the neutron star is balanced by the ram
pressure of the accreting matter) which, for typical values of the
magnetic field and the mass accretion rate, can be quite close (a few
neutron star radii) to the compact object.  In this situation, the
neutron star can be accelerated by the accretion of matter and angular
momentum from a (Keplerian) accretion disk to very short periods, in
principle up to the limiting period (usually of the order of or below
1 ms), which depends on the mass-radius relation of the neutron star,
and therefore on the equation of state of ultra-dense matter. At the
end of the mass transfer phase, these systems will be observed as low
magnetized, very fast (millisecond) pulsars in a binary system with a
very low mass (if any) companion star; these systems are indeed
observed in radio and form the class of MSP.

This evolutionary scenario was spectacularly confirmed by the
discovery of millisecond coherent pulsations in LMXBs; this important
discovery arrived recently, in 1998, when coherent millisecond
pulsations with a period of 2.5 ms where discovered in the transient
LMXB \saxj\ \citep{wij98}, thanks to the large
effective area ($\sim 6000$ cm$^2$) and high time resolution (up to 1
$\mu$sec) of the Proportional Counter Array (PCA) on board the {\it
Rossi} X-ray Timing Explorer (RXTE). \saxj, for the rest a quite
common LMXB, belongs to a close binary system, $P_{\rm orb} \simeq
2$~h \citep{cha98}, and is the only one among the known
accreting millisecond pulsars which has shown more than one X-ray
outburst in the RXTE era. We now know seven accreting millisecond
pulsars (see \citep{wij05}, \citep{gal06a}  for reviews); all of
them are X-ray transients in very compact systems (orbital period
between 40 min and 4 h), the fastest of which is \igr, with a spin
period of $P_{\rm spin} \simeq 1.7$~ms, and the slowest of which is
\0929, with a spin period of $P_{\rm spin} \simeq 5.4$~ms.

In this paper we review recent results from timing analysis of a
sample of accreting millisecond pulsars. In particular, we first
describe the timing technique we have used to analyse these sources,
and then we present the results we have obtained to date for four (out
of seven) accreting millisecond pulsars.

\section{Timing Analysis}

The timing analysis we have applied to our sample of accreting
millisecond pulsars is based on standard timing techniques that are
fully described in \citep{bur07}, and that we briefly describe
here.

We use the fact that orbital periods of these systems (typically less
than a few hours) are much smaller than the timescale on which
intrinsic variations of the spin period are expected to produce their
effects on the pulse phase delays (typically tens of days). Also,
variations of the phase delays caused by uncertainties in the source
position on the sky are expected to show their effects on even longer
timescales (a fraction of a year). We therefore proceed in the
following way. First of all, the arrival times of all the events are
reported to the Solar system barycenter, adopting the best estimate of
the source position on the sky. These arrival times are then corrected
for the Doppler effects caused by the binary motion using the best
estimates of the orbital parameters and using the following
approximation:
\begin{equation}
\label{eq1}
t_{em} \simeq t_{arr} - x \sin \left[\frac{2 \pi}{P_{orb}} 
(t_{arr} - T^*) \right],
\end{equation}
where $x = a \sin i / c$ is the projected semi-major axis in
light-seconds, $T^*$ is the time of the ascending node passage at the
beginning of the observation, and $t_{em}$ and $t_{arr}$ are the
emission and the arrival times, respectively. From the corrected time
series we calculate the pulse phase delays fitting the pulse profiles,
obtained from time intervals of length of the order of one orbital 
period, with one or more sinusoids (according to the harmonic content 
of the pulse profile) with fixed periods, and we plot the phase of the 
fundamental (and, when present, the phase of the harmonics) as a function
of time to study the phase delays evolution.

The differential of expression~(\ref{eq1}) with respect to the orbital
parameters allows to calculate the uncertainties, $\sigma_{\phi, orb}$, 
on the phase delays induced by the uncertainties on the orbital parameters. 
Possible errors in the adopted values of the orbital parameters will
indeed show up as a timing noise in the pulse phase delays of amplitude
$\sigma_{\phi, orb}$, and we sum in quadrature these uncertainties to
the corresponding statistical uncertainties found from the fit of the
pulse profile. 
This noise can be further reduced calculating the phase delays over intervals 
of length of the order of one orbital period, since the uncertainties induced 
by the errors on the orbital parameters are sinusoids of period $P_{orb}$.

On the other hand, possible errors on the assumed position of the 
source in the sky will give a modulation of the phase delays with
a period of 1 year. Since the X-ray observations span time intervals
of a few tens of days at most (pulsations from accreting millisecond
pulsars are usually observed during X-ray outbursts which last typically
less than a month), this induced modulation on the phase delays will
appear like a linear trend or a parabola (the first orders of the
series expansion of sinusoids with period 1 year). Therefore,
the uncertainties in the source position can be considered like a
systematic error on the linear correction to the spin period (linear
term) and/or on the spin period derivative (parabolic term).
This means that it is mandatory to have a good knowledge of the source
position on the sky (such that is derived from optical or radio 
counterparts) in order to reduce these systematics.

In all our timing analyses we have included all these possible source
of error, considering the uncertainties induced by errors in the orbital
parameters as an additive noise in the pulse phase delays, and
the uncertainties induced by errors in the source position as systematic
uncertainties in the period and its derivative. In all the cases 
reported below, the systematics resulted much smaller than (or at most 
of the same order of magnitude of) the statistical errors on the linear
and quadratic terms derived from the fit of the phase delays.

\subsection{\igr}

\igr\ was discovered by the {\it INTEGRAL} satellite in December 2005,
when it showed an X-ray outburst which lasted from December 3 to 21.
X-ray pulsations were significantly detected only during the first 12 days
of the outburst.
With a spin period of $1.7$ ms is the fastest among the known accreting
millisecond pulsars and belongs to a binary system with orbital period 
$\sim 2.5$ h \citep{gal05}. Falanga et al.\ \citep{fal05} report for this 
source the presence of a constant spin-up of $\sim 8 \times 10^{-13}$ Hz/s.
Burderi et al.\ \citep{bur07}  re-analysed these data and, in particular, fitted
the phase delays vs.\ time with physical models taking into account the
observed decrease of the X-ray flux as a function of time during the X-ray
outburst, in order to get a valuable estimate of the mass accretion rate
onto the compact object. In fact, in the hypothesis the the spin-up of the
source is caused by the accretion of matter and angular momentum from a 
Keplerian accretion disk, the mass accretion rate, $\dot M$ onto the neutron 
star can be calculated by the simple relation $2 \pi I \dot \nu = \dot M 
(G M R)^{1/2}$, where $I$ is the moment of inertia of the neutron star, 
$\dot \nu$ the spin frequency derivative, $G$ the Gravitational constant, 
$M$ the mass of the compact object, $R$ the accretion radius, and 
$(G M R)^{1/2}$ the Keplerian specific angular momentum at the accretion 
radius. Since we are neglecting any threading effect of the magnetic field
in the accretion disk outside the accretion radius, the estimate of 
$\dot M$ derived in this way should be considered as a lower limit.

\begin{figure}
\label{figigr}
  \includegraphics[height=.4\textheight]{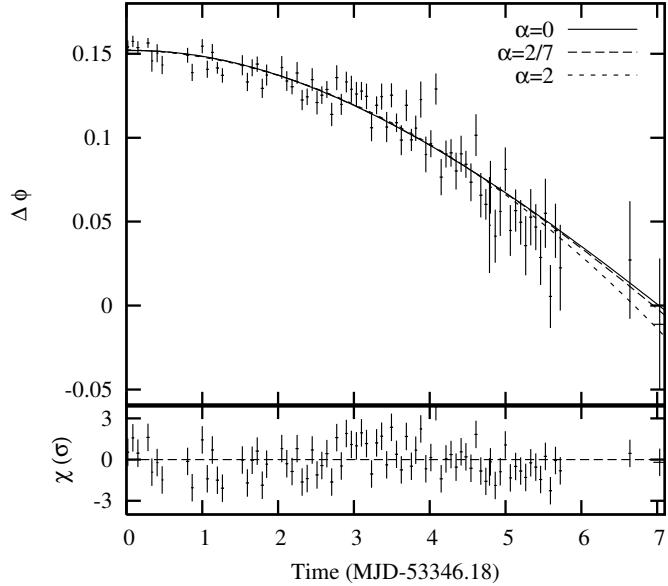} 
  \caption{Pulse phases computed folding at the spin period reported in
Table~\ref{tab1} and plotted versus time together with the best fit
curves (upper panel) and residuals in units of $\sigma$ with respect to
the model with $\alpha = 2/7$ (lower panel).}
\end{figure}

Because the X-ray flux, which is assumed to be a good tracer of the mass
accretion rate, is observed to decrease along the outburst, this has to be
included in the relation in order to obtain the correct value of the 
mass accretion rate at the beginning of the outburst as well as its 
temporal evolution (note that the accretion radius also depends on the
mass accretion rate, $R \propto \dot M^\alpha$ where $\alpha$ is usually
assumed to be 2/7, and therefore is a function of time). 
From an analysis of the X-ray light curve, we concluded that the flux
was, in good approximation, decreasing linearly with time, and adopted
the following dependence: $\dot M (t) = \dot M_0 [1 - (t-T_0)/t_B]$,
where $T_0$ is the time at the beginning of the observation (Dec 7), 
and $t_B = 8.4$ days. We can therefore derive the following expression
for the expected evolution of the phase delays vs.\ time:
\begin{equation}
\label{eq2}
\phi = - \phi_0 - \Delta \nu_0\, (t - T_0) -
\frac{1}{2} \dot \nu_{0} (t - T_0)^2
\left[ 1 - \frac{(2-\alpha)(t-T_0)}{6 t_B} \right] ,
\end{equation}
where $\phi_0$ is a constant, $\Delta \nu_0$ is the linear correction to
the value of the spin frequency adopted to produce the pulse profiles, 
and $\dot \nu_{0}$ is the frequency derivative at $t = T_0$.
For the fit we used three possible values for $\alpha$, i.e.\ 
a) the standard $\alpha = 2/7$ which corresponds to assuming that the 
accretion radius is proportional to the Alfven radius, b) $\alpha = 0$
which corresponds to an accretion radius equal to the corotation radius
(the radius at which the Keplerian frequency equals the neutron star 
spin frequency, that is the maximum radius at which accretion can occur),
and c) $\alpha = 2$ which corresponds to a simple parabolic function, that is
to a constant mass accretion rate. For all the assumed values of $\alpha$
we obtained acceptable fits, and we have calculated the lower limit to the
mass accretion rate at the beginning of the outburst, that is obtained in 
the case $\alpha = 0$: 
\begin{equation}
\label{dotm}
\dot M_{-10} = 5.9 \times \dot \nu_{-13}\, I_{45}\, m^{-2/3},
\end{equation}
where $\dot M_{-10}$ is $\dot M_0$ in units of $10^{-10}\, 
M_\odot$ yr$^{-1}$, $\dot \nu_{-13}$ is $\dot \nu_0$ in units of 
$10^{-13}$ s$^{-2}$, $I_{45}$ is $I$ in units of $10^{45}$ g cm$^2$,
and $m$ is the mass of the neutron star in units of \msun. 
We adopt the FPS equation of state for the neutron star matter for 
$m = 1.4$ and the spin frequency of \igr\ which gives $I_{45}= 1.29$ and
$R_{NS}= 1.14 \times 10^6$ cm (see e.g.\ \citep{coo94}).
Pulse phase delays and the corresponding best fit functions for all the
three values of $\alpha$ are shown in Fig.~\ref{figigr}.

From the fitting of the phase delays with these relations we 
find $\dot \nu_{-13} = 11.7$, and a lower limit to the mass accretion rate 
of $\dot M_{-10} \sim 70 \pm 10$ (case $\alpha = 0$). This would correspond
to a bolometric luminosity of $\sim 7 \times 10^{37}$ ergs/s. This is about
an order of magnitude higher than the X-ray luminosity inferred from the
observed X-ray flux and assuming a distance of 5 kpc.
Burderi et al.\ \citep{bur07} have argued that, since the pulse profile is very
sinusoidal with negligible harmonic content, we probably just see only
one of the two emitting polar caps, and therefore the observer intercepts 
just half of the total emitted X-ray luminosity. In this way, we can reduce 
the discrepancy between the bolometric luminosity inferred from the mass
accretion rate and the observed X-ray luminosity, but still we need to 
place the source to a quite large distance of $7.4 - 10.7$ kpc (note that
10 kpc is close to the edge of the Galaxy in the direction if \igr).
Other possible explanations for the discrepancy between the mass accretion
rate inferred by the timing and the observed X-ray luminosity can be 
that the energy released by accretion is not completely converted into
X-ray luminosity, but a fraction of this is released in other energy bands
or other emission mechanisms, or that the moment of inertia is smaller than
the assumed value. As soon as we will have a direct, independent, estimate of 
the distance to the source, we will have very important information on the
$\dot M$ - X-ray luminosity relation and/or on the physical parameters of 
the neutron star.

\subsection{\saxj}

\begin{figure}
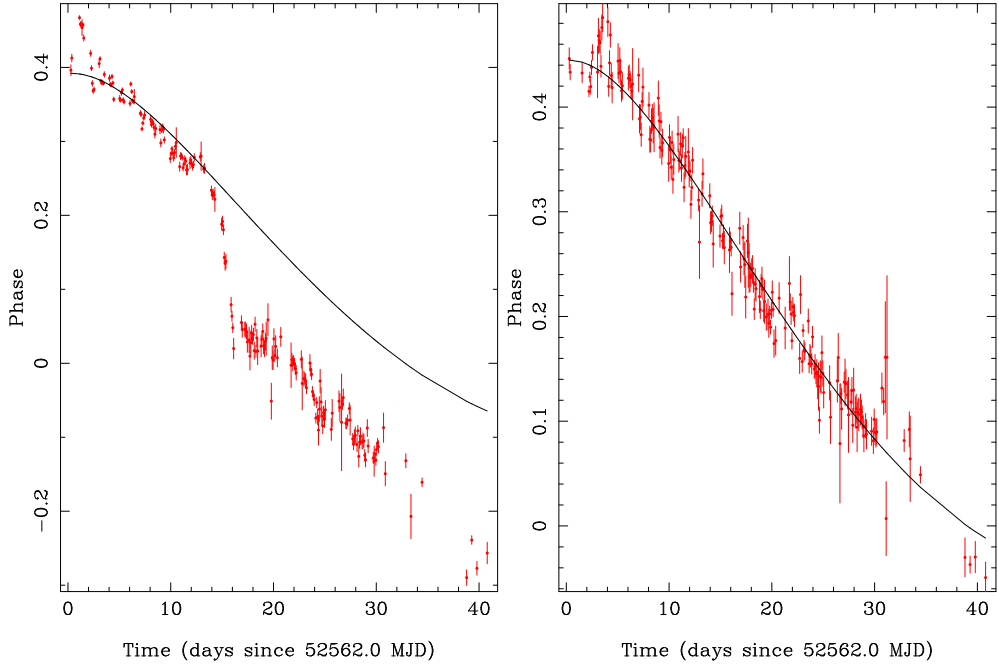

\label{figsaxj}
  \includegraphics[height=.4\textheight]{f2a_saxj.eps}
  \includegraphics[height=.4\textheight]{f2b_saxj.eps}
  \caption{{\bf Left:} Phase vs.\ time for the fundamental of the pulse frequency
    of \saxj. {\bf Right:} Phase vs.\ time for the first harmonic of the pulse
    frequency of \saxj. On top of the data, the best fit function (the sum of
    eq.~(\ref{phit}) and a quadratic term) is plotted as a solid line.}
\end{figure}

\saxj\ is the only one among accreting millisecond pulsars to have shown
more than one X-ray outburst during the RXTE era; it goes into
outburst roughly every two years (in 1998, 2000, 2002, and 2005 up to
date). We have performed a timing analysis of the 2002 outburst, which
lasted about 40 days from October 15 to November 26, one of the most
extensively covered by RXTE observations (see details in \citep{bur06}). In this case, the pulse profile shows the presence of a
significant first harmonic, and we therefore studied both the phase
delays of the fundamental and the phase delays of the harmonic as a
function of time. If we look at the phase delays derived from the
fundamental in Fig.~\ref{figsaxj}, we can see a very puzzling
behavior, since a rather fast phase shift is present at day 14 from
the beginning of the outburst. Interestingly, as it can be easily seen
from the X-ray light curve of the outburst, day 14 corresponds to a
change in the steepness of the exponential decay with time of the
X-ray flux.  On the other hand, the phase delays of the harmonic do
not show any evidence of this phase jump. This is not an effect of the
worse statistics we have for the phase delays derived from the
harmonic, which of course show larger error bars. We have to conclude
that the phase jump in the fundamental is not related to a intrinsic
spin variation (which would have affected the whole pulse profile),
but is instead caused by a change of the shape of the pulse profile
(probably caused by the same mechanism causing the increase of the
steepness of the exponential decay of the X-ray flux).

Given the regular behavior of the phase delays of the harmonic, we tried
to fit these to an appropriate model. As in the case of \igr, we considered
a varying with time mass accretion rate, that in this case is exponentially 
decreasing during the outburst: $\dot M (t) = \dot M_0 \exp(t - T_0) / \tau$,
where $\tau = 9.27$ days can be derived from a fit of the (first 14 days)
light curve. Again we can derive the expected variation of the phase delays
for the case of an exponentially decreasing mass accretion rate:
\begin{equation}
\label{phit}
\phi (t) = \phi_0 - B (t - T_0) - C \exp(t - T_0) / \tau,
\end{equation}
where $C = 1.067 \times 10^{-4} I_{45}^{-1} P_{-3}^{1/3} m^{2/3} \tau^2
\dot M_{-10}$, $P_{-3}$ is the spin period in millisecond, and 
$B = \Delta \nu_0 + C / \tau$. 
However, we obtained a poor fit both using the expression above or using
a simple parabolic trend. Indeed, with the model of eq.~\ref{phit} we can
obtain a good fit of the first 14 days of the outburst, but, with respect
to this fit, we observe a flattening of the phase delays after day 14.
To describe this flattening we therefore added to eq.~\ref{phit} a 
quadratic term corresponding to a constant spin-down. The best fit
function is plotted on top of the data in Fig.~\ref{figsaxj}.

From this best fit we can derive a spin-up at the beginning of the
outburst of $\dot \nu_0 \sim 4.4 \times 10^{-13}$ Hz/s, corresponding
to a mass accretion rate of $\dot M_{-10} = 18$, and a (marginally
significant\footnote{Due to the uncertainty in the modelization of the
X-ray flux behavior vs.\ time in the second half of the outburst}) 
constant spin-down of $\dot \nu_{sd} \sim -7.6 \times 10^{-14}$ Hz/s.  
In the case of \saxj\ the distance to the source is known and is about 3.5
kpc \citep{gal06}; therefore we can check if also in this
case there is a discrepancy between the mass accretion rate inferred
from the timing results and the observed X-ray luminosity.  We still
find a discrepancy, but, in this case, the mass accretion rate inferred
from timing is only a factor of 2 larger than the X-ray luminosity, since 
this is about $1 \times 10^{37}$ ergs/s (see \citep{bur06}).  The 
spin-down observed at the end of the outburst can be explained, for instance, 
by a threading of the accretion disk by the neutron star magnetic field
outside the accretion radius. Of course, in agreement with what we
observe, we expect that such a threading effect will be observable at
the end of the outburst, when the mass accretion rate significantly
decreases  (see \citep{rap04}). We can therefore evaluate
the magnetic moment, $\mu$, of \saxj\ from our measured value of the 
spin-down, using the relation 
$\mu^2 / (9 R_{CO}^3) = 2 \pi I \dot \nu_{sd}$,
where $R_{CO}$ is the corotation radius. The magnetic field found
in this way is $B \sim (3.5 \pm 0.5) \times 10^8$ Gauss, perfectly
in agreement with previous constraints \citep{dis03}.

\subsection{\1814}

\1814 was discovered in 2003 by RXTE \citep{mar03}; the X-ray
outburst started on June 5 and lasted about 53 days. The spin period
is $\sim 3.14$ ms and the binary orbital period, $P_{orb} = 4.275$ h, 
is the largest in the sample of known accreting millisecond pulsars.
We have performed a timing analysis of this source (see \citep{pap07}, and the paper by Papitto et al.\ included in this volume) finding 
that the neutron star shows a global spin-down, 
$\dot \nu_{sd} \sim (-6.7 \pm 0.7) \times 10^{-14}$ Hz/s, 
during all the outburst.
Again this source shows a puzzling behavior of the phase delays; for
this source the harmonic contain in the pulse profile is quite high.
In particular we detect the fundamental and the first harmonic, whose
amplitudes are both quite large. We have therefore plotted the phase 
delays of both the fundamental and the harmonic, finding in this case
that both show the same trend when plotted vs.\ time. This trend is
approximately parabolic and showing a global spin-down of the pulsar,
but superposed to this general trend we find oscillations of the phase
delays.

Differently from previous cases, the X-ray flux in this source does not
monotonically decay during the outburst; instead the flux is observed to
oscillate around a mean value during the first 30 days of the outburst,
and then it fastly decays to quiescence. We find that the oscillations 
observed in the phase delays of the fundamental and the harmonic are very
well anticorrelated with the oscillations present in the X-ray flux.
We have therefore interpreted, similarly to the case of the phase shift 
observed in \saxj, these oscillations as phase shifts induced by small
movements of the magnetic field footpoints in the neutron star surface
driven by variations of the X-ray flux.

Also in this case, we can try to get an estimate of the neutron star 
magnetic field from the observed global spin-down trend and using the
threading of the accretion disk model (as in \citep{rap04}).
We get a quite large value for the magnetic field of 
$\sim 8 \times 10^8$ Gauss.\footnote{The magnetic field has been 
evaluated using for the average mass accretion rate during the first
35 days of the outburst, when the mass accretion rate can be considered
almost constant, the value $\sim 5.4 \times 10^{-10}$ \msun/yr, and
assuming a distance to the source of 8 kpc.}

\subsection{\0929}

\0929 is a high-latitude source and was discovered by RXTE in 2002, when 
it showed an X-ray outburst which started on May 2 and lasted for about 
53 days.  Galloway et al.\ \citep{gal02} provided an orbital solution for this 
source, reporting a quite short period of $\sim 44$ min. With a spin 
period of $\sim 5.4$ ms this is the slowest among the known sample of 
accreting millisecond pulsars. They also performed a timing analysis of 
the pulse phase delays, showing that the source underwent a steady 
spin-down while accreting of $\dot \nu = -9.2(4) \times 10^{-14}$ Hz/s.
We have re-analysed these data, using an improved source position on the 
sky, and basically confirm the already reported results, although with
a revised spin-down rate of $\dot \nu = -5.5(4) \times 10^{-14}$ Hz/s.

Although the timing results seem quite simple in this case, indeed \0929
shows the most puzzling behavior with respect all the sample of accreting
millisecond pulsars discussed here. In fact, as in the case of \igr,
\0929 shows an almost linear decrease of the X-ray flux during the outburst,
with a decay time $t_B \simeq 58.5$ days. This means that, along the outburst,
the expected spin-up should decrease, and the global derivative of the 
spin frequency (that is the sum of the spin-up and spin-down rates) should
show an increasing global spin-down along the outburst. However, the pulse
phase delays do not show any increasing spin-down, since the best fit 
suggests a constant (or at most decreasing) spin-down. If the decreasing 
of the X-ray flux does not affect the behavior of the phase delays, this 
means that the corresponding spin-up rate should be always negligible
with respct to the observed spin-down, i.e.\ 
$\dot \nu_{su} << -\dot \nu_{sd} \sim  5.5 \times 10^{-14}$ Hz/s. 
If we assume, in agreement with the request above, that the spin-up is 
at least a factor of 5 lower than the spin-down rate, we find a 
corresponding mass accretion rate of $\dot M < 6 \times 10^{-11}$ \msun/yr,
which would correspond to a quite low bolometric luminosity of 
$< 6 \times 10^{35}$ ergs/s. If we compare this luminosity with the
observed X-ray luminosity, $L_X \sim 1.0 \times 10^{37}$ d$_{5\;kpc}^2$
ergs/s where d$_{5\;kpc}^2$ is the distance to the source in units of
5~kpc, we find an upper limit to the source distance  of about 
1.2~kpc, that is less than the lower limit of 5~kpc derived by the 
expected secular mass accretion rate (driven by gravitational radiation)
and a supposed outburst recurrence time $> 6.5$ yr \citep{gal02}.

A distance of $\sim 1$~kpc is unlikely to be correct, although we have to 
note that \0929 is a high latitude source, and therefore the closer is the 
source, the smaller will be the height of the source above the Galactic 
plane. Otherwise, the reason of this discrepancy may be in the used model
for the threading (spin-down) torque; in most of the models this depends 
only on the magnetic field strength and should therefore remain constant 
along an X-ray outburst. However, the pulse phase delays seem to suggest
that the spin-down in \0929 may decrease at the end of the outburst. 
We note that results of MHD simulations on the interaction between the 
accretion disc and the magnetosphere of a NS in the propeller regime 
(i.e. when the accretion radius is larger than corotation radius) presented
by \citep{rom04}; see also  \citep{rom05}) show that the
spin down torque resulting from this interaction may decrease with decreasing
accretion rate, with the material torque owing to the accreted
matter relegated to a marginal role in building the overall torque. 
This behavior is related to the weaker coupling between the
magnetosphere and the disc matter corresponding to a lower accretion rate,
which has the effect to weaken the toroidal component of
the magnetic field in the magnetosphere, which is the one responsible for
the spinning down of the pulsar.
We therefore need a correct modelization of the spin-down torque in order
to correctly evaluate the upper limit on the mass accretion rate from timing.

\section{Discussion and Conclusions}  

In this paper we review the results of a timing analysis performed over a sample
of accreting millisecond pulsars (a summary of these results is shown in 
Tab.~\ref{tab1}).
\begin{table}
\begin{tabular}{lrrr}
\hline \tablehead{1}{l}{b}{Source}
  & \tablehead{1}{c}{b}{P$_{\rm orb}$ (h)}
  & \tablehead{1}{c}{b}{P$_{\rm spin}$ (ms)}
  & \tablehead{1}{c}{b}{$\dot \nu$ (Hz/s)}   \\
\hline
\igr  & $2.456692\;(2)$ & $1.66974977466\;(5)$ & $1.2\;(2) \times 10^{-12}$   \\
\saxj & $2.01365469\;(3)$ & $2.4939197632\;(4)$ & $4.4\;(8) \times 10^{-13}$  \\
\saxj &    &  		     & $-7.6\; (1.5) \times 10^{-14}$  \\
\1814 & $4.27464525\;(5)$  & $3.1811056697\;(1)$ & $-6.7\;(7) \times 10^{-14}$  \\
\0929 & $0.7263183\;(8)$ & $5.4023317862\;(3)$ & $-5.5\;(4) \times 10^{-14}$  \\
\hline
\end{tabular}
\caption{Summary of timing results for our sample of accreting millisecond pulsars}
\label{tab1}
\end{table}
We have showed that a few accreting millisecond pulsars, which are
supposed to accrete from a Keplerian accretion disk, show steady 
spin-down while accreting. The only (thus far) available explanation
for this is in terms of the magnetic field - accretion disk interaction,
that is a threading of the accretion disk by the magnetic field outside
the accretion radius. However, this predicts a quite low luminosity in 
the case of \0929, and therefore a quite small distance to the source.
Independent measurements of the distance to \0929 will give important 
information on the torque acting on the neutron star and its response.
Most of our results are puzzling but many of them are exactly as expected:
\igr\ shows the strongest spin-up, in agreement with the fact that it is
the fastest accreting millisecond pulsars; slower pulsars show less spin-up
or spin-down. In \saxj\ we observe a shift in phase of the fundamental
at day 14, the same day at which is observed a steepness of the exponential
decay of the X-ray flux, and again around that day there seems to be a change 
from spin-up to spin-down of the pulsar. These facts are in agreement with a
scenario where some sort of ejection mechanism becomes important in the 
disk when the mass accretion rate is sufficiently low; this explains the
increased steepness in the flux decay and the possible change from global 
spin-up to global spin-down. This may also be responsible of movements of
the footpoints of the magnetic field onto the neutron star surface, and
therefore of the change of the shape of the pulse profile, that is observed 
as a shift in phase of the fundamental, although the detailed mechanism 
is not clear yet.


\begin{theacknowledgments}
This work was supported by the Ministero della Istruzione, della Universit\`a e 
della Ricerca (MIUR), national program PRIN2005 2005024090\_004.

\end{theacknowledgments}

\end{document}